\documentclass{article}
\usepackage[utf8]{inputenc}
\usepackage{color}

\title{Threshold Voltage Control in Dual-Gate Organic Electrochemical Transistors}
\author{Hsin Tseng$^{*1}$, Anton Weissbach$^1$, Juzef Kucinski$^1$, Ali Solgi$^1$, \\Rakesh Nair$^1$, Lukas M Bongartz$^1$, Giuseppe Ciccone$^1$, Matteo Cucchi$^{1,\dagger}$, \\Karl Leo$^1$, \& Hans Kleemann$^{*1}$\\\\\small{$^1$Dresden Integrated Center for Applied Physics and}\\ \small{Photonic Materials (IAPP) and Institute for Applied Physics,}\\\small{Technische Universität Dresden, Nöthnitzer Str. 61,}\\ \small{01187 Dresden, Germany}\\\small{$\dagger$ Present address: Laboratory for Soft Bioelectronic Interfaces Neuro-X Institute,}\\\small{Ecole Polytechnique Fédérale de Lausanne (EPFL),}\\\small{Geneva, Switzerland}}

\usepackage[numbers,sort&compress]{natbib}
\usepackage{graphicx}

\begin{document}

\maketitle

\section*{Abstract}
Organic electrochemical transistors (OECTs) based on Poly(3,4-ethylenedioxythiophene):poly(styrene sulfonic acid) (PEDOT:PSS) are a benchmark system in organic bioelectronics. In particular, the superior mechanical properties and the ionic-electronic transduction yield excellent potential for the field of implantable or wearable sensing technology. However, depletion-mode operation PEDOT:PSS-based OECTs cause high static power dissipation in electronic circuits, limiting their application in electronic systems. Hence, having control over the threshold voltage is of utmost technological importance. Here we demonstrate PEDOT:PSS-based dual-gate OECTs with solid-state
electrolyte where the threshold voltage is seamlessly adjustable during operation. We show that the degree of threshold voltage tuning linearly depends
on the gate capacitance, which is a straightforward approach for circuit designers to adjust the threshold voltage only by the device dimensions. The
PEDOT:PSS-based dual-gate OECTs show excellent device performance and can be pushed to accumulation-mode operation, resulting in a simplified and relaxed design of complementary inverters.

\section{Introduction}
Organic electrochemical transistors(OECTs) have recently been in the spotlight of research because of their use in bioelectronics \cite{rivnay2018organic, rashid2021organic, someya2016rise}, neuromorphic computing \cite{cucchi2021reservoir, van2018organic, harikesh2022organic, krauhausen2021organic}, and biological or chemical sensors \cite{tseng2021membrane,romele2020multiscale, liu2021ultrafast,guo2021rapid}. All these application scenarios are based on the conduction mechanism in organic mixed ionic-electronic conductors (OMIECs) \cite{paulsen2020organic}, enabling efficient translation of ionic or chemical signals into electronic signals and vice versa. In addition, OMIECs offer easy production by printing and excellent mechanical properties \cite{liao2015flexible, gualandi2016textile}. With the development of solid-state electrolytes \cite{khodagholy2012organic, weissbach2022photopatternable, hemantha2022ionic, jo2018gelatin}, OECTs might be integrated into wearable or even implantable systems with intelligent sensor function.\\\\
In a common OECT, ions from the electrolyte penetrate the OMIEC polymer matrix and the distribution of ions in the OMIEC can be manipulated applying a bias voltage to the gate electrode. As the concentration of mobile holes / electrons in the OMIEC is regulated by the ion concentration via an electrochemical redox reaction, the gate bias can be used to control the conductance of the transistor channel and hence switch the transistor on and off \cite{cucchi2022thermodynamics}. 
\\\\
One important device parameter of OECTs for sensing, computing, and circuitry is the threshold voltage ($\mathrm{V_{th}}$). It is the point when the gate bias switches the transistors between high current accumulation regime and low current depletion regime. It has a great technological relevance for example for logic gates where it determines the trip-point of digital inverters. Furthermore, the threshold voltage is also of great importance for sensing applications as in the sub-threshold regime, the current through the transistor exponentially depends on the gate bias offering the highest possible sensitivity of the system (comes with the drawback of losing linearity of the sensor). Hence, having control over the threshold voltage is of utmost technological importance and several strategies have been proposed to tune the threshold voltage by adjusting device parameters or materials. For example, changing the gate electrode material affects the potential drop at the gate electrode and changes the OECT operation regime from capacitive to Faradaic \cite{tarabella2010effect}, leading to different threshold voltages. Alternatively, modifying the gate electrode with redox-active species or dopants controls the gate's work function \cite{doris2018dynamic, tan2022tuning}, also serving the same purpose. In addition, the design of the channel material, such as its chemical structure, influences the interactions with electrolytes and can, therefore, regulate OECTs operation and performance \cite{zeglio2018active, keene2020enhancement}. However, even without manipulating the material system, the threshold voltage can be tuned solely by the device geometry. For example, the gating efficiency of the same gate electrode material can be improved as the gate capacitance is increased \cite{koutsouras2021efficient}, and thus reducing the threshold voltage. 
\\\\
The issue of threshold control is particularly relevant for the most often used OECT material: Poly(3,4-ethylenedioxythiophene) polystyrene sulfonate (PEDOT:PSS) is a working horse OECTs material because it is easy to process, commercially available, and shows high transconductance. PEDOT:PSS is, however, highly doped, resulting in depletion-mode (normally-on) transistor behavior. A depletion-mode device is unfavorable for circuitry because of high power dissipation. In particular, for the design of inverter circuits, which are a basic element of digital electronics, the depletion-mode behavior is very disadvantageous as it complicates the inverter design and reduces the noise margin and gain of the circuits. Unfortunately, chemically modifying the gate electrode does not give an accumulation-mode device, although it helps tuning the threshold voltage \cite{doris2018dynamic, tan2022tuning}. On top of that, using this approach, the threshold voltage cannot be adjusted anymore once the device has been manufactured. Although the chemical modification of either the gate electrode or channel material offers control over the threshold voltage, controlling it via the device design is technological preferable as the use of different materials, e.g., for the gate and the channel or additional chemical doping significantly increases the complexity of device and circuit fabrication. 
\\\\
Using a dual-gate architecture is an alternative strategy to control the threshold voltage by the device design. This approach has been put forth for conventional organic thin-film transistors and precise control over the threshold voltage as well as tunability during operation have been demonstrated \cite{spijkman2011dual,guo2020vertical, Guo2021}. Two gate insulators are used to isolate the semiconductor channel from the top and bottom gate electrodes in these architectures. Thereby, two channels are formed at opposing interfaces of the semiconductor layer, which are used to regulate the conductance of the transistor. Having separated electrolytes, this design principle could also be adopted to OECTs using gate electrodes at the bottom and top. However, OECTs are usually fabricated in a side-gate configuration due to the conductive nature of the electrolyte. This geometry is advantageous because of its easy processing and high production yield. In contrast to a typical dual-gate thin-film transistor \cite{spijkman2011dual}, the two gates in a dual-gate OECT in side-gate configuration would be in a shared electrolyte and it is not clear whether the two gates can be used to control the threshold voltage independently or they simultaneously influence the device performance. Recently, Ji et al. demonstrated a dual-liquid-gated OECT using electropolymerization to modify the gate capacitance with PEDOT:PSS \cite{ji2021dual} and showed that the transconductance can be tuned to some extent. However, accurate control over the threshold voltage was not possible with their approach, and most importantly, they could not make PEDOT:PSS-based OECTs operate as accumulation-mode transistors. 
\\\\
Here, we demonstrate PEDOT:PSS-based dual-gate OECTs with solid-state electrolyte where the threshold voltage can be continuously tuned during operation. We show that the degree of tuning the threshold voltage linearly depends on the gate capacitance which is a straight-forward approach for circuit designers to adjust the threshold voltage only by the device dimensions. The PEDOT:PSS-based dual-gate OECTs, which can be densely integrated using conventional photolithography or printing techniques, show excellent device performance and can be pushed to accumulation-mode operation, leading to simplified processing, relaxed design requirements, and improved performance of complementary inverters.

\section{Results and Discussion}
\begin{figure}[!]
    	\centering
    	\includegraphics[width=\linewidth]{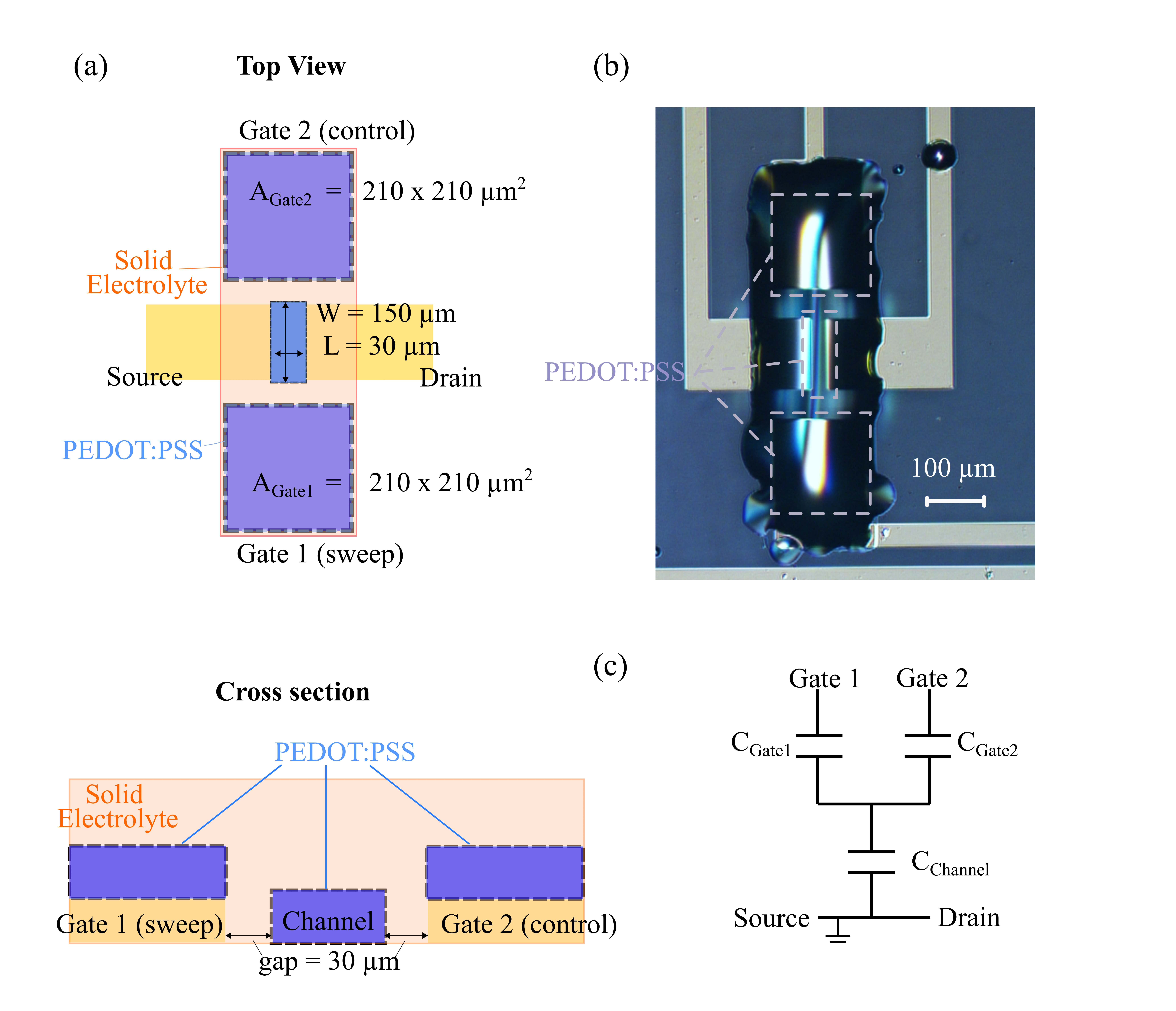}
    	\caption{(a) Top view and cross section configuration of a PEDOT:PSS-based dual-gate OECT, including source and drain electrodes, PEDOT:PSS channel, and two PEDOT:PSS gates symmetric with respect to the channel. All the PEDOT:PSS films have the same thickness of 100 nm.} (b) Optical microscopic image of a dual-gate OECT. (c) Equivalent circuit model of the dual-gate OECT. 
    
    	\label{fig:configuration}
    \end{figure}

\begin{figure}[!]
    	\centering
    	\includegraphics[width=\linewidth]{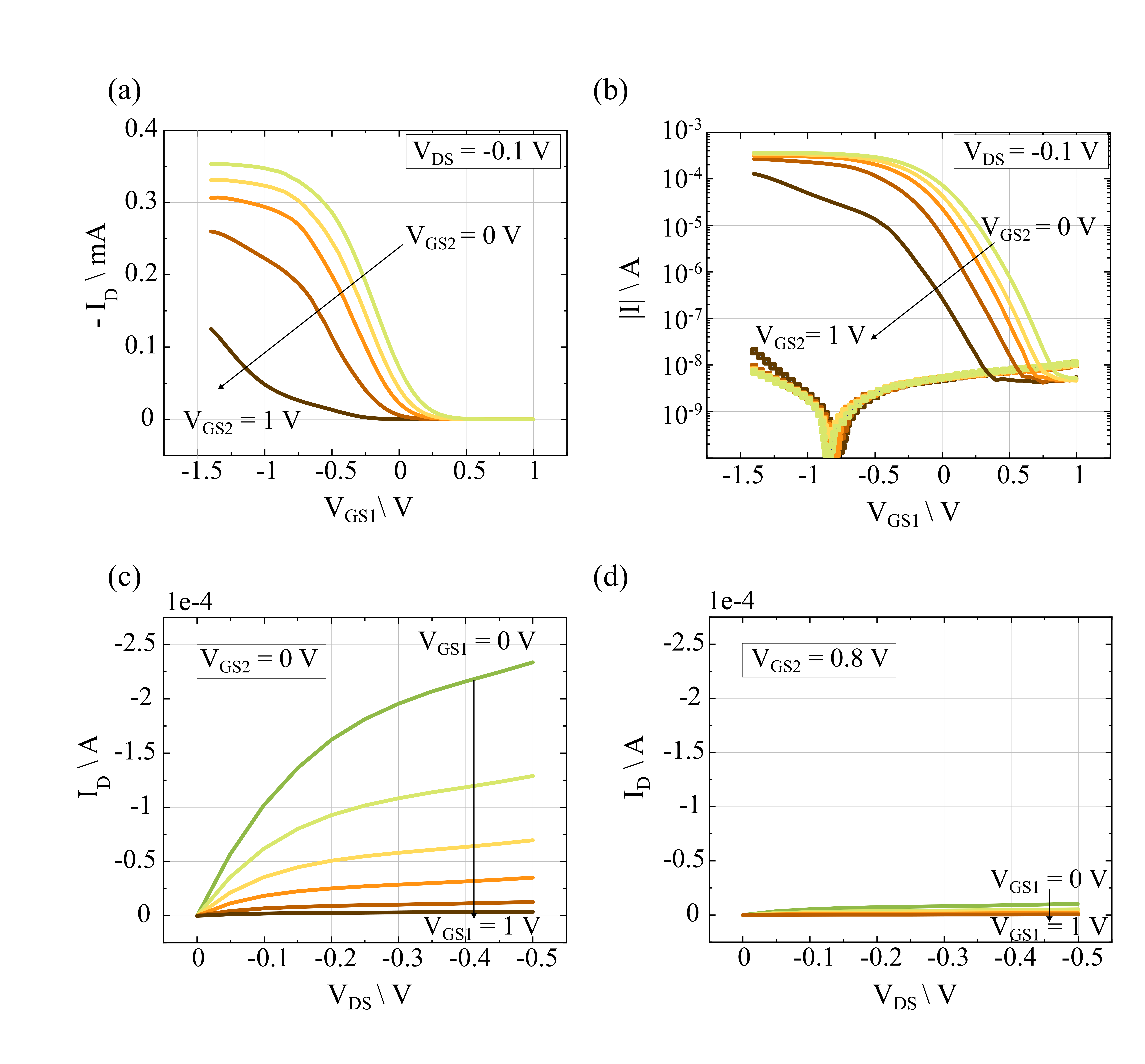}
    	\caption{Electrical characterization of the PEDOT:PSS-based dual-gate OECT with $\mathrm{A_{Gate 2} = A_{Gate 1} = 44100}$ $\mathrm { \mu m^{2}}$. (a) Transfer characteristic in linear scale at $\mathrm{V_{DS}}$ = -0.1 V. $\mathrm{V_{GS2}}$ is fixed for the loop of $\mathrm{V_{GS1}}$ sweeping from -1.5 V to +1 V. (b) Transfer characteristic in logarithmic scale, including the drain current (solid line) and the gate current (dotted line). The same color of the drain current and the gate current means they are under the same $\mathrm{V_{GS2}}$. (c) Output characteristic of the dual-gate OECT at $\mathrm{V_{GS2}}$ = 0 V. (d) Output characteristic at $\mathrm{V_{GS2}}$ = 0.8 V.
    	}
    	\label{fig:transfer}
    \end{figure}

\begin{figure}[!]
    	\centering
    	\includegraphics[width=\linewidth]{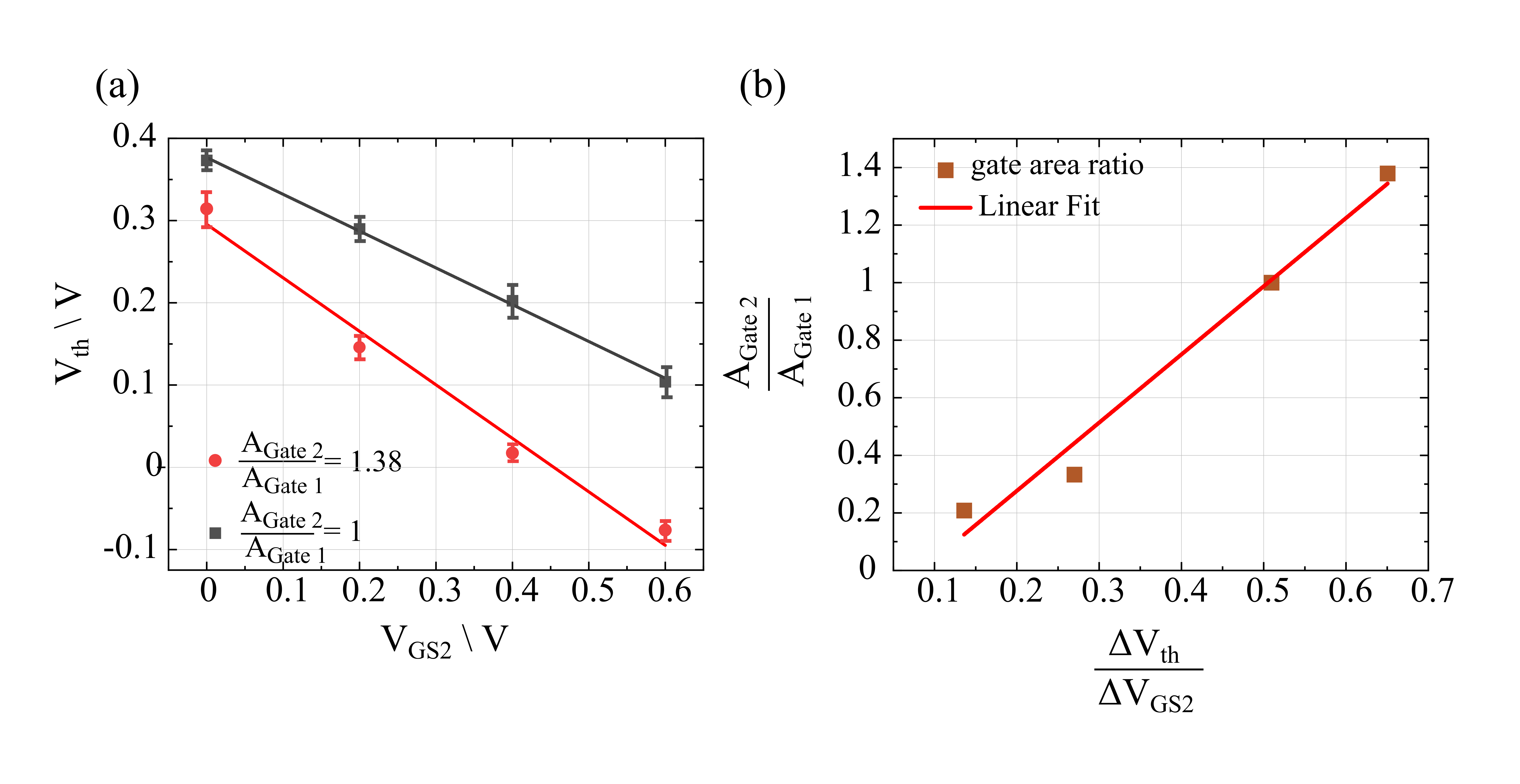}
    	\caption{(a) Threshold voltage of a PEDOT:PSS-based dual-gate OECT as a function of $\mathrm{V_{GS2}}$ for different gate area ratios. The larger area of Gate 2 (red) gives a steeper slope, namely a larger degree of tuning. Each data point is a mean value for 5 devices and each device is measured 3 times. (b) The degree of threshold voltage tuning increases with the ratio of the gate area.
    	}
    	\label{fig:slope}
    \end{figure}

\begin{figure}[!]
    	\centering
    	\includegraphics[width=\linewidth]{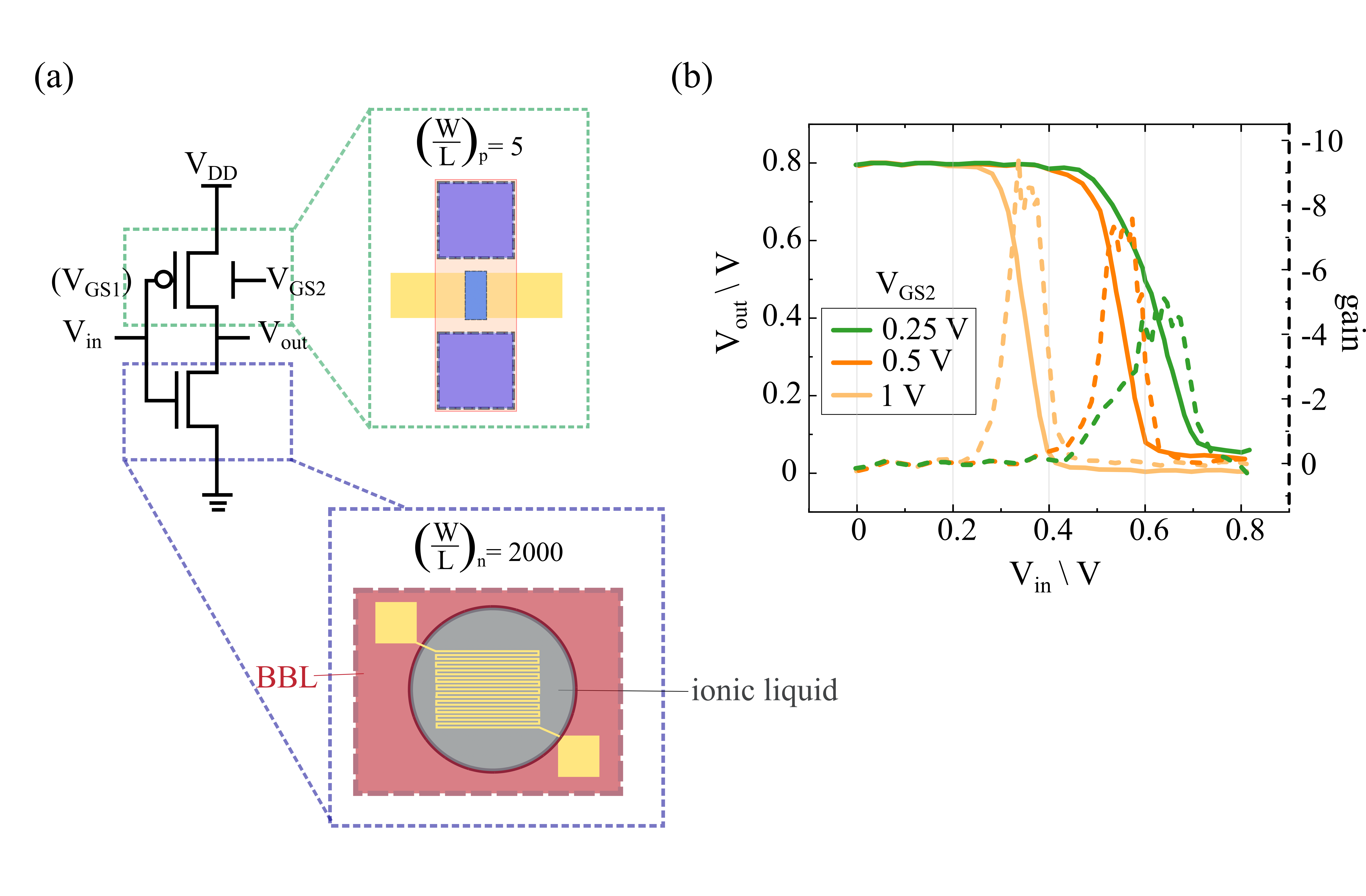}
    	\caption{(a) A complementary inverter layout of a PEDOT:PSS-based dual-gate OECT with $(\frac{W}{L})_p = 5$ and a BBL OECT with $(\frac{W}{L})_n = 2000$. The input voltage $\mathrm{V_{in}}$ applies to the Gate 1 of the PEDOT:PSS-based dual-gate OECT and to the gate of the BBL OECT. $\mathrm{V_{GS2}}$ is constantly applied at the Gate 2 of the PEDOT:PSS-based dual-gate OECT. The device configuration of both types of OECTs; the gate electrode of BBL OECT is Ag/AgCl immersed in the ionic liquid. Further device dimension can be found in the Experimental section. (b) Inverter transfer characteristic: as the threshold voltage of the PEDOT:PSS-based dual-gate OECT is tuned by increasing $\mathrm{V_{GS2}}$ (from 0.25 V, 0.5 V to 1 V), the transfer curve (solid line) shifts to the left, and the inverter gain (dashed line) increases.
    	}
    	\label{fig:inverter}
    \end{figure}

Figure \ref{fig:configuration}(a, b) show the schematic layout of a dual-gate OECT. The transistor consists of two in-plane gate electrodes, a sweeping gate (Gate 1) and a controlling gate (Gate 2), the semiconductor channel with source and drain electrodes, and the solid-state electrolyte. PEDOT:PSS is used as the semiconductor channel material as well as for both gate electrodes in order to increase the capacitance of the gate. Using the same material for the gate and the channel reduces the fabrication complexity, and the volumetric capacitance of a PEDOT:PSS-based gate strongly reduces the voltage loss at the gate/electrolyte interface \cite{koutsouras2021efficient}. 
\\\\
The capacitance of the PEDOT:PSS-based gate can be scaled by film area \cite{bianchi2020scaling, volkov2017understanding, rivnay2015high} on the condition that this film is formed at the same spin-coating process as the channel material, giving the same thickness of 100 nm. To evaluate the effect of the capacitance of the control gate (Gate 2) on the threshold voltage, we only vary the area of Gate 2, from 9200 $\mathrm{\mu m^{2}}$, 14700 $\mathrm{\mu m^{2}}$, 44100 $\mathrm{\mu m^{2}}$, to 60900 $\mathrm{\mu m^{2}}$, leaving the area of both channel and Gate 1 and the distance of 30 $\mathrm{\mu m}$ between the gate and the channel fixed (cf. Figure \ref{fig:configuration}). The solid-state electrolyte is inkjet-printed on top of the channel and the gate electrodes, followed by UV-light induced cross-linking \cite{weissbach2022photopatternable}. More details on the fabrication process of these integrated OECTs are given in the Experimental Section.
\\\\
Figure \ref{fig:transfer} presents the electrical characterization of a dual-gate OECT with $\mathrm{A_{Gate 2} = A_{Gate 1} = 44100}$ $\mathrm{\mu m^{2}}$. The transfer characteristics are measured at a drain-source bias $\mathrm{V_{DS}}$ of -0.1 V for Gate 2 bias ranging from 0 V to +1 V in steps of 0.2 V. The transfer curves systematically shift with the applied $\mathrm{V_{GS2}}$. For a controlling bias $\mathrm{V_{GS2}}$ of 0 V (grounded), the transfer curve at the far right in Figure \ref{fig:transfer} (a, b), the effect of the Gate 2 is negligible. The dual-gate OECT behaves like a single-gate OECT and only Gate 1 redistributes ionic charges in the electrolyte. 
\\\\
The effect of the $\mathrm{V_{GS2}}$ $>$ 0 V on the dual-gate OECT is shown in the curves on the left in Figure \ref{fig:transfer} (a, b). Polarons in PEDOT:PSS are neutralized by cations driven by $\mathrm{V_{GS2}}$. To compensate for $\mathrm{V_{GS2}}$ and achieve the original drain current at $\mathrm{V_{GS2}}$ of 0 V, the sweeping gate $\mathrm{V_{GS1}}$ has to be increased to more negative values. Therefore, the transfer curve and hence the threshold voltage is shifted to the left. The interplay between the bias on the two gate electrodes determines the current in the dual-gate OECT at a given drain-source bias. The gate current in Figure \ref{fig:transfer} (b) is significantly lower than the drain current because of the precise patterning technology of the solid-state electrolyte.
\\\\
As shown in Figure \ref{fig:transfer}(c), output curves are measured at $\mathrm{V_{GS2} = }$0 V. The drain current is only affected by $\mathrm{V_{GS1}}$. Further, in Figure \ref{fig:transfer}(d), the drain current is simultaneously influenced by $\mathrm{V_{GS1}}$ and $\mathrm{V_{GS2}}$; at the same scale of x-axis and y-axis, the drain current is close to 0 A, proving that the PEDOT-PSS-based OECT in fact operates as an accumulation-mode transistor (in agreement with the transfer curve shown in Figure \ref{fig:transfer}(a)). It is worth to mention that the effect of inhomogenous dedoping becomes relevant at high $\mathrm{V_{DS}}$ where the curve is supposed to saturate \cite{kaphle2020finding}. In this study, we report on the change of threshold voltage for small $\mathrm{V_{DS}}$=-0.1 V where inhomoegenous dedoping can be ignored and the threshold voltage is well defined. If inhomoegenous dedoping comes into play, the threshold voltage becomes a function of the drain-source voltage. Still, the $\mathrm{V_{GS2}}$ can be used to manipulate the transconductance of the transistor.
\\\\
We postulate that the dual-gate OECT device behavior can be modelled as two parallel gate capacitor connected in series to the channel capacitor, as shown in Figure \ref{fig:configuration}(c). This is because the electrolyte resistance (200k$\Omega$) is negligible compared to the shunt resistance of the electrochemical double layers formed at the semiconductor-electrolyte interface (typical leakage current in the range of 10 nA at 1 V as shown in Figure \ref{fig:transfer} (b)). The function of the solid-state electrolyte does not differ from that of a liquid electrolyte such as $\mathrm{NaCl_{(aq)}}$. The solid polymer structure of the solid-state electrolyte forms a matrix for the movement of the ionic liquid. As water has been used as a solvent for the solid-state electrolyte, the PEDOT:PSS layer is always in a swollen state, which allows ions from the ionic liquid to move in and out of the PEDOT:PSS layer and thus dopes and dedopes the semiconductor \cite{weissbach2022photopatternable}. Accordingly, the voltage only drops across the gate capacitance/ channel capacitance and the solid-state electrolyte can be treated as an equipotential surface. 
The total effective gate-source voltage influencing the channel is then determined by the total capacitance of $\mathrm{C_{Gate 1}}$ and $\mathrm{C_{Gate 2}}$ and the voltages applied. 
\\\\
We extract the threshold voltage ($\mathrm{V_{th}}$) from the transfer characteristics in Figure \ref{fig:transfer}(a). The $\mathrm{V_{th}}$ is defined by plotting the drain current against the gate-source voltage, linearly fitting this curve, and intercepting the value on the x-axis of $\mathrm{V_{GS1}}$. The $\mathrm{V_{DS}}$ of -0.1 V is chosen to extract $\mathrm{V_{th}}$ is to avoid the effect of non-uniform dedoping in our system \cite{kaphle2020finding}. Figure \ref{fig:slope}(a) presents the threshold voltage as a function of the controlling gate $\mathrm{V_{GS2}}$. The data is in a mean value for 5 devices for each geometry and clearly shows the shift in threshold voltage. It should be noted that these solid-state electrolyte OECTs show a significant hysteresis in the transfer curve \cite{weissbach2022photopatternable}, which makes it technically speaking impossible to derive a single threshold voltage value. However, using the dual-gate configuration, we observe that the transfer curve including hysteresis is homogeneously shifted. Hence, for a sake of simplicity, we only plot the transfer curve for switching the device from on- to off-state thereby ignoring the hysteresis.
\\\\
When the area of Gate 2 is enlarged, the slope in Figure \ref{fig:slope}(a) increases, which allows us to turn these PEDOT:PSS-based OECTs from depletion- to accumulation-mode operation. Figure \ref{fig:slope}(b) shows that the degree of controlling the threshold voltage in dual-gate OECTs linearly scales with the ratio of the gate area (being equal to the ratio of capacitances). Using the equivalent circuit proposed in Figure \ref{fig:configuration}(c), we can predict the scaling of the threshold voltage shift as a function of the gate area ratio by the following expression:
\begin{equation}
    \mathrm{V_{th}} = \frac{\mathrm{V_{th}^0} (\mathrm{A_{Gate 1}+A_{Gate 2}+A_{Channel})}}{\mathrm{A_{Gate 1}}}-\frac{\mathrm{V_{GS2}A_{Gate 2}}}{\mathrm{A_{Gate 1}}}
\end{equation}
where the constant $\mathrm{V_{th}^0}$ represents the threshold voltage without the presence of Gate 2. As shown in Figure \ref{fig:slope}(b), the experimental data fit very well to the model predictions and even without over-sizing Gate 2 significantly, a strong tunability of $\mathrm{V_{th}}$ is achieved. In fact, accumulation-mode operation can be already achieved if the area of Gate 2 is only 1.38-times larger than the area of Gate 1 (Figure \ref{fig:slope}(a)).
\\\\
We demonstrate the advantage of this dual-gate OECT technology for logic circuits. As an example, a complementary inverter, combining a p-type and an n-type OECT, is chosen here as the most simple logic circuit. It works as a digital amplifier and is often combined with OECT-based sensors to increase the biosignal sensitivity for bioelectronics \cite{romele2020multiscale}.
The most common used n-type semiconductor material for OECTs is poly(benzimidazobenzophenanthroline) (BBL) \cite{zeglio2018active,sun2018complementary, sun2018n, jia2019emerging}, with which an OECT operates in an accumulation-mode. Due to the depletion-mode operation of PEDOT:PSS-based OECTs and the large transconductance compared to BBL-based devices, good inverter characteristics can only be achieved if the BBL-based OECT is significantly larger than the PEDOT:PSS-based device. The channel width to length ratio of the BBL-based devices is typically chosen to be at least thousand times larger than the ratio of the PEDOT:PSS-based device (e.g., 16000-times larger in Ref.\cite{romele2020multiscale}). 
\\\\A complementary inverter layout with a PEDOT:PSS-based dual-gate OECT (p-type) and a BBL-based OECT (n-type) is shown in Figure \ref{fig:inverter}(a). The input voltage ($\mathrm{V_{in}}$) is applied to the common gate, i.e., the gate of the BBL device and the Gate 1 of the PEDOT:PSS-based dual-gate OECT and is swept from 0 V to 0.8 V. The supply voltage $\mathrm{V_{DD}}$ is set to 0.8 V. A constant voltage at Gate 2 ($\mathrm{V_{GS2} = }$0.25 V, 0.5 V, 1 V) is applied during the inverter measurement to control the threshold voltage of the p-channel device. The output voltage ($\mathrm{V_{out}}$) is measured to determine the transfer curve of the inverter which is shown in Figure \ref{fig:inverter}(b). As $\mathrm{V_{GS2}}$ increases, the inverter transfer curve shifts to the left. In particular, the trip point of the inverter is seamlessly adjusted by $\mathrm{V_{GS2}}$ as it controls the threshold voltage of the dual-gate devices. The dual-gate design of OECTs offers a more robust design and operation of circuits, which can be used to improve the sensitivity of any bioelectronics system and thus contributes greatly to the field of bioelectronics.
\\\\
%\textcolor{red}{The dual-gate OECTs shows great potential contributions to bioelectronics. It can be exploited for a more robust design and operation of circuits. The dual-gate device enables to shift the trip point and adjust the gain in a complementary inverter. An inverter, working as a digital amplifier, is often combined with OECT-based sensors to increase the biosignal sensitivity OECT-based sensors are often combined with a digital amplifier, e.g. an inverter, in order to increase the sensitivity of the system \cite{romele2020multiscale}. Using The dual-gate design of OECTs can be exploited for a more robust design and operation of circuits. It is of great help to the field of bioelectronics as OECT-based sensors are often combined with a digital amplifier, e.g. an inverter, in order to increase the sensitivity of the system}

\clearpage

\vfill

\clearpage

\vfill
\clearpage

\vfill
\clearpage
\vfill
\clearpage

\section{Conclusion}

In conclusion, we demonstrate continuous tuning of the threshold voltage in PEDOT:PSS-based dual-gate OECTs. These dual-gate structures are easy to fabricate, employing the often used side-gate architecture. The threshold voltage scales linearly with the voltage at the control gate (Gate 2), and the degree of tuning linearly increases with the gate area ratio. Furthermore, we modeled the device behavior with an equivalent circuit, and the experimental data fit very well with the model predictions. The
PEDOT:PSS-based dual-gate OECTs, which can be densely integrated using conventional photolithography or printing techniques, show excellent device performance, and they can be pushed to accumulation-mode operation, leading to
improved performance and relaxed design requirements of complementary OECT inverters. 

\section{Experimental}
\textit{Device fabrication}: 
The process of structuring the electrode and channel pattern of a dual-gate OECT follows ref.\cite{tseng2021membrane}. Source, drain, and gate electrodes were patterned on a glass substrate with 50 nm Au and 3 nm Cr by photolithography and wet-etching using Standard Gold Etchant and Standard Chromium Etchant. PEDOT:PSS-based solution was prepared with 95 wt.\% of PEDOT:PSS (Hearaeus Clevios PH 1000, 1.1 wt.\% solids in water, 1:2.5) and 5 wt.\% of ethylene glycol. This solution was spin-coated at 3000 rpm on the electrodes and the 100 nm-PEDOT:PSS thin film was patterned by fluorine-based photolithography \cite{kleemann2012direct} and dry etching \cite{hoppner2020precise} using O$_{2}$ and Ar. The solid-state electrolyte precursor solution containing 1 mL deionized water, 750 mg N-isopropylacrylamide, 20 mg N,N'-methylenebisacrylamide, 200 mg 2-hydroxy-4'-(2-hydroxyethoxy)-2-methylpropiophenone, and 1.5 mL 1-ethyl-3-methylimidazolium ethyl sulfate \cite{weissbach2022photopatternable} was inkjet printed on top of the active area followed by 2 minutes UV cross-linking. The PEDOT:PSS-based dual-gate OECT was stored in the glovebox overnight and then was encapsulated with glass for further measurement.

BBL solution was prepared by dissolving 5 mg BBL (Sigma Aldrich) in 1 mL methanesulfonic acid and stirring at 70 $^\circ$C overnight. The BBL solution was then spin-coated at 1000 rpm on a gold substrate with W=10 mm, L=5 $\mathrm{\mu m}$. Afterwards, the BBL film was soaked into ethanol for 1 minute and then dried on a hot plate at 150 $^{o}$C for 5 minutes, and the resulting BBL film is 70 nm. \\\\
\textit{Electrical characteristics}: Transfer and output characterizations were done with Keithley SMUs controlled by the software SweepMe! BBL OECTs were measured with a Ag/AgCl gate and ionic liquid  1-ethyl-3-methylimidazolium ethyl sulfate in an ambient condition, giving $\mathrm{V_{th}}$ = 0.11 V and $\mathrm{g_{m}}$ = 0.73 mS.

\bibliographystyle{unsrtnat} %按照出現順序引用

\begin{thebibliography}{10}

\bibitem{rivnay2018organic}
Jonathan Rivnay, Sahika Inal, Alberto Salleo, R{\'o}is{\'\i}n~M Owens, Magnus
  Berggren, and George~G Malliaras.
\newblock Organic electrochemical transistors.
\newblock {\em Nature Reviews Materials}, 3(2):1--14, 2018.

\bibitem{rashid2021organic}
Reem~B Rashid, Xudong Ji, and Jonathan Rivnay.
\newblock Organic electrochemical transistors in bioelectronic circuits.
\newblock {\em Biosensors and Bioelectronics}, 190:113461, 2021.

\bibitem{someya2016rise}
Takao Someya, Zhenan Bao, and George~G Malliaras.
\newblock The rise of plastic bioelectronics.
\newblock {\em Nature}, 540(7633):379--385, 2016.

\bibitem{cucchi2021reservoir}
Matteo Cucchi, Christopher Gruener, Lautaro Petrauskas, Peter Steiner, Hsin
  Tseng, Axel Fischer, Bogdan Penkovsky, Christian Matthus, Peter Birkholz,
  Hans Kleemann, and Karl Leo.
\newblock Reservoir computing with biocompatible organic electrochemical
  networks for brain-inspired biosignal classification.
\newblock {\em Science Advances}, 7(34):eabh0693, 2021.

\bibitem{van2018organic}
Yoeri van De~Burgt, Armantas Melianas, Scott~Tom Keene, George Malliaras, and
  Alberto Salleo.
\newblock Organic electronics for neuromorphic computing.
\newblock {\em Nature Electronics}, 1(7):386--397, 2018.

\bibitem{harikesh2022organic}
Padinhare~Cholakkal Harikesh, Chi-Yuan Yang, Deyu Tu, Jennifer~Y Gerasimov,
  Abdul~Manan Dar, Adam Armada-Moreira, Matteo Massetti, Renee Kroon, David
  Bliman, Roger Olsson, Eleni Stavrinidou, Magnus Berggren, and Simone Fabiano.
\newblock Organic electrochemical neurons and synapses with ion mediated
  spiking.
\newblock {\em Nature Communications}, 13(1):1--9, 2022.

\bibitem{krauhausen2021organic}
Imke Krauhausen, Dimitrios~A Koutsouras, Armantas Melianas, Scott~T Keene,
  Katharina Lieberth, Hadrien Ledanseur, Rajendar Sheelamanthula, Alexander
  Giovannitti, Fabrizio Torricelli, Iain Mcculloch, Yoeri van De~Burgt, and
  Paschalis Gkoupidenis.
\newblock Organic neuromorphic electronics for sensorimotor integration and
  learning in robotics.
\newblock {\em Science Advances}, 7(50):eabl5068, 2021.

\bibitem{tseng2021membrane}
Hsin Tseng, Matteo Cucchi, Anton Weissbach, Karl Leo, and Hans Kleemann.
\newblock Membrane-free, selective ion sensing by combining organic
  electrochemical transistors and impedance analysis of ionic diffusion.
\newblock {\em ACS Applied Electronic Materials}, 3(9):3898--3903, 2021.

\bibitem{romele2020multiscale}
Paolo Romele, Paschalis Gkoupidenis, Dimitrios~A Koutsouras, Katharina
  Lieberth, Zsolt~M Kov{\'a}cs-Vajna, Paul~WM Blom, and Fabrizio Torricelli.
\newblock Multiscale real time and high sensitivity ion detection with
  complementary organic electrochemical transistors amplifier.
\newblock {\em Nature Communications}, 11(1):1--11, 2020.

\bibitem{liu2021ultrafast}
Hong Liu, Anneng Yang, Jiajun Song, Naixiang Wang, Puiyiu Lam, Yuenling Li,
  Helen Ka-wai Law, and Feng Yan.
\newblock Ultrafast, sensitive, and portable detection of covid-19 igg using
  flexible organic electrochemical transistors.
\newblock {\em Science Advances}, 7(38):eabg8387, 2021.

\bibitem{guo2021rapid}
Keying Guo, Shofarul Wustoni, Anil Koklu, Escarlet D{\'\i}az-Galicia,
  Maximilian Moser, Adel Hama, Ahmed~A Alqahtani, Adeel~Nazir Ahmad,
  Fatimah~Saeed Alhamlan, Muhammad Shuaib, Arnab Pain, Iain McCulloch, Stefan~T
  Arold, Raik Gr{\"u}nberg, and Sahika Inal.
\newblock Rapid single-molecule detection of covid-19 and mers antigens via
  nanobody-functionalized organic electrochemical transistors.
\newblock {\em Nature Biomedical Engineering}, 5(7):666--677, 2021.

\bibitem{paulsen2020organic}
Bryan~D Paulsen, Klas Tybrandt, Eleni Stavrinidou, and Jonathan Rivnay.
\newblock Organic mixed ionic--electronic conductors.
\newblock {\em Nature Materials}, 19(1):13--26, 2020.

\bibitem{liao2015flexible}
Caizhi Liao, Meng Zhang, Mei~Yu Yao, Tao Hua, Li~Li, and Feng Yan.
\newblock Flexible organic electronics in biology: materials and devices.
\newblock {\em Advanced materials}, 27(46):7493--7527, 2015.

\bibitem{gualandi2016textile}
Isacco Gualandi, Marco Marzocchi, Andrea Achilli, D~Cavedale, Annalisa
  Bonfiglio, and Beatrice Fraboni.
\newblock Textile organic electrochemical transistors as a platform for
  wearable biosensors.
\newblock {\em Scientific Reports}, 6(1):1--10, 2016.

\bibitem{khodagholy2012organic}
Dion Khodagholy, Vincenzo~F Curto, Kevin~J Fraser, Moshe Gurfinkel, Robert
  Byrne, Dermot Diamond, George~G Malliaras, Fernando Benito-Lopez, and
  Roisin~M Owens.
\newblock Organic electrochemical transistor incorporating an ionogel as a
  solid state electrolyte for lactate sensing.
\newblock {\em Journal of Materials Chemistry}, 22(10):4440--4443, 2012.

\bibitem{weissbach2022photopatternable}
Anton Weissbach, Lukas~M Bongartz, Matteo Cucchi, Hsin Tseng, Karl Leo, and
  Hans Kleemann.
\newblock Photopatternable solid electrolyte for integrable organic
  electrochemical transistors: operation and hysteresis.
\newblock {\em Journal of Materials Chemistry C}, 10:2656--2662, 2022.

\bibitem{hemantha2022ionic}
CP~Hemantha~Rajapaksha, Pushpa~Raj Paudel, PM~Sineth~G Kodikara, Drona Dahal,
  Thiloka~M Dassanayake, Vikash Kaphle, Bj{\"o}rn L{\"u}ssem, and Antal
  J{\'a}kli.
\newblock Ionic liquid crystal elastomers-based flexible organic
  electrochemical transistors: Effect of director alignment of the solid
  electrolyte.
\newblock {\em Applied Physics Reviews}, 9(1):011415, 2022.

\bibitem{jo2018gelatin}
Young~Jin Jo, Ki~Yoon Kwon, Zia~Ullah Khan, Xavier Crispin, and Tae-il Kim.
\newblock Gelatin hydrogel-based organic electrochemical transistors and their
  integrated logic circuits.
\newblock {\em ACS applied materials \& interfaces}, 10(45):39083--39090, 2018.

\bibitem{cucchi2022thermodynamics}
Matteo Cucchi, Anton Weissbach, Lukas~M Bongartz, Richard Kantelberg, Hsin
  Tseng, Hans Kleemann, and Karl Leo.
\newblock Thermodynamics of organic electrochemical transistors.
\newblock {\em Nature Communications}, 13(1):1--8, 2022.

\bibitem{tarabella2010effect}
Giuseppe Tarabella, Clara Santato, Sang~Yoon Yang, Salvatore Iannotta, George~G
  Malliaras, and Fabio Cicoira.
\newblock Effect of the gate electrode on the response of organic
  electrochemical transistors.
\newblock {\em Applied Physics Letters}, 97(12):205, 2010.

\bibitem{doris2018dynamic}
Sean~E Doris, Adrien Pierre, and Robert~A Street.
\newblock Dynamic and tunable threshold voltage in organic electrochemical
  transistors.
\newblock {\em Advanced Materials}, 30(15):1706757, 2018.

\bibitem{tan2022tuning}
Siew Ting~Melissa Tan, Gijun Lee, Ilaria Denti, Garrett LeCroy, Kalee
  Rozylowicz, Adam Marks, Sophie Griggs, Iain McCulloch, Alexander Giovannitti,
  and Alberto Salleo.
\newblock Tuning organic electrochemical transistor threshold voltage using
  chemically doped polymer gates.
\newblock {\em Advanced Materials}, page 2202359, 2022.

\bibitem{zeglio2018active}
Erica Zeglio and Olle Ingan{\"a}s.
\newblock Active materials for organic electrochemical transistors.
\newblock {\em Advanced Materials}, 30(44):1800941, 2018.

\bibitem{keene2020enhancement}
Scott~T Keene, Tom~PA van~der Pol, Dante Zakhidov, Christ~HL Weijtens,
  Ren{\'e}~AJ Janssen, Alberto Salleo, and Yoeri van~de Burgt.
\newblock Enhancement-mode pedot: Pss organic electrochemical transistors using
  molecular de-doping.
\newblock {\em Advanced Materials}, 32(19):2000270, 2020.

\bibitem{koutsouras2021efficient}
Dimitrios~A Koutsouras, Fabrizio Torricelli, Paschalis Gkoupidenis, and Paul~WM
  Blom.
\newblock Efficient gating of organic electrochemical transistors with in-plane
  gate electrodes.
\newblock {\em Advanced Materials Technologies}, 6(12):2100732, 2021.

\bibitem{spijkman2011dual}
Mark-Jan Spijkman, Kris Myny, Edsger~CP Smits, Paul Heremans, Paul~WM Blom, and
  Dago~M De~Leeuw.
\newblock Dual-gate thin-film transistors, integrated circuits and sensors.
\newblock {\em Advanced Materials}, 23(29):3231--3242, 2011.

\bibitem{guo2020vertical}
Erjuan Guo, Zhongbin Wu, Ghader Darbandy, Shen Xing, Shu-Jen Wang, Alexander
  Tahn, Michael G{\"o}bel, Alexander Kloes, Karl Leo, and Hans Kleemann.
\newblock Vertical organic permeable dual-base transistors for logic circuits.
\newblock {\em Nature Communications}, 11(1):1--9, 2020.

\bibitem{Guo2021}
Erjuan Guo, Shen Xing, Felix Dollinger, Ren{\'e} H{\"u}bner, Shu-Jen Wang,
  Zhongbin Wu, Karl Leo, and Hans Kleemann.
\newblock Integrated complementary inverters and ring oscillators based on
  vertical-channel dual-base organic thin-film transistors.
\newblock {\em Nature Electronics}, 4(8):588--594, Aug 2021.

\bibitem{ji2021dual}
Jianlong Ji, Hongwang Wang, Ran Liu, Xiaoning Jiang, Qiang Zhang, Yubo Peng,
  Shengbo Sang, Qijun Sun, and Zhong~Lin Wang.
\newblock Dual-liquid-gated electrochemical transistor and its neuromorphic
  behaviors.
\newblock {\em Nano Energy}, 87:106116, 2021.

\bibitem{bianchi2020scaling}
Michele Bianchi, Stefano Carli, Michele Di~Lauro, Mirko Prato, Mauro Murgia,
  Luciano Fadiga, and Fabio Biscarini.
\newblock Scaling of capacitance of pedot: Pss: volume vs. area.
\newblock {\em Journal of Materials Chemistry C}, 8(32):11252--11262, 2020.

\bibitem{volkov2017understanding}
Anton~V Volkov, Kosala Wijeratne, Evangelia Mitraka, Ujwala Ail, Dan Zhao, Klas
  Tybrandt, Jens~Wenzel Andreasen, Magnus Berggren, Xavier Crispin, and Igor~V
  Zozoulenko.
\newblock Understanding the capacitance of pedot: Pss.
\newblock {\em Advanced Functional Materials}, 27(28):1700329, 2017.

\bibitem{rivnay2015high}
Jonathan Rivnay, Pierre Leleux, Marc Ferro, Michele Sessolo, Adam Williamson,
  Dimitrios~A Koutsouras, Dion Khodagholy, Marc Ramuz, Xenofon Strakosas,
  Roisin~M Owens, Christian Benar, Jean-Michel Badier, Christophe Bernard, and
  George~G Malliaras.
\newblock High-performance transistors for bioelectronics through tuning of
  channel thickness.
\newblock {\em Science Advances}, 1(4):e1400251, 2015.

\bibitem{kaphle2020finding}
Vikash Kaphle, Pushpa~Raj Paudel, Drona Dahal, Raj~Kishen Radha~Krishnan, and
  Bj{\"o}rn L{\"u}ssem.
\newblock Finding the equilibrium of organic electrochemical transistors.
\newblock {\em Nature communications}, 11(1):1--11, 2020.

\bibitem{sun2018complementary}
Hengda Sun, Mikhail Vagin, Suhao Wang, Xavier Crispin, Robert Forchheimer,
  Magnus Berggren, and Simone Fabiano.
\newblock Complementary logic circuits based on high-performance n-type organic
  electrochemical transistors.
\newblock {\em Advanced Materials}, 30(9):1704916, 2018.

\bibitem{sun2018n}
Hengda Sun, Jennifer Gerasimov, Magnus Berggren, and Simone Fabiano.
\newblock n-type organic electrochemical transistors: materials and challenges.
\newblock {\em Journal of Materials Chemistry C}, 6(44):11778--11784, 2018.

\bibitem{jia2019emerging}
Hanyu Jia and Ting Lei.
\newblock Emerging research directions for n-type conjugated polymers.
\newblock {\em Journal of Materials Chemistry C}, 7(41):12809--12821, 2019.

\bibitem{kleemann2012direct}
Hans Kleemann, Alex~A Zakhidov, Merve Anderson, Torben Menke, Karl Leo, and
  Bj{\"o}rn L{\"u}ssem.
\newblock Direct structuring of c60 thin film transistors by photo-lithography
  under ambient conditions.
\newblock {\em Organic Electronics}, 13(3):506--513, 2012.

\bibitem{hoppner2020precise}
Marco H{\"o}ppner, David Kneppe, Hans Kleemann, and Karl Leo.
\newblock Precise patterning of organic semiconductors by reactive ion etching.
\newblock {\em Organic Electronics}, 76:105357, 2020.

\end{thebibliography}

\section*{Acknowledgements}

We thank the European Social Fund (Project OrgaNanoMorph, project number: 100382168) and the Hector Fellow Academy (30000619) for providing the financial support. Furthermore, the authors thank the Bundesministerium für Bildung und Forschung (BMBF) for funding from the project BAYOEN (01IS21089).
The EMERGE project has received funding from the European Union’s Horizon 2020 research and Innovation programme under grant agreement Nº 101008701.
\section*{Data Availability}
The data that support the findings of this study are available from the corresponding author upon reasonable request.

\end{document}